\documentclass[twocolumn,aps,prl,groupedaddress, nobalancelastpage]{revtex4}
\usepackage{graphicx}
\usepackage{color}
\usepackage{amsmath}
\usepackage{marvosym}
\usepackage{bm}
\usepackage{verbatim}

\usepackage{amssymb}
\usepackage{amsthm}
\usepackage{amsfonts}
\usepackage{bbm}
\usepackage{hyperref}
\usepackage{stmaryrd}

\newcommand{\be}{\begin{equation}}
\newcommand{\ee}{\end{equation}}
\newcommand{\bea}{\begin{eqnarray}}
\newcommand{\eea}{\end{eqnarray}}

\newcommand{\ra}{\rangle}

\newcommand{\Ket}[1]{| #1\ra}

\renewcommand{\vec}[1]{{\bf #1}}

\newcommand{\dbar}{\raisebox{-0.1ex}[\height][0pt]{$\mathchar'26$}\mkern-12mu {d}}

\begin{document}

\title{Quantum plasmonic non-reciprocity in parity-violating magnets} 

\author{Arpit Arora$^1$}
\author{Mark S. Rudner$^2$}
\author{Justin C. W. Song$^{1}$}
\affiliation{$^1$Division of Physics and Applied Physics, School of Physical and Mathematical Sciences, Nanyang Technological University, Singapore 637371}
\affiliation{$^2$Department of Physics, University of Washington, Seattle WA 98195, USA}

\begin{abstract}
The optical responses of metals are often dominated by plasmonic resonances -- the collective oscillations of interacting electron liquids. Here we unveil a new class of plasmons -- quantum metric plasmons (QMPs) -- that arise in a wide range of parity violating magnetic metals. In these materials, a dipolar distribution of the quantum metric (a fundamental characteristic of Bloch wavefunctions) produces intrinsic non-reciprocal bulk plasmons. Strikingly, QMP non-reciprocity manifests even when the single-particle dispersion is symmetric: QMPs are sensitive to time-reversal and parity violations hidden in the Bloch wavefunction. In materials with asymmetric single-particle dispersions, quantum metric dipole induced non-reciprocity can continue to dominate at large frequencies. We anticipate that QMPs can be realized in a wide range of parity violating magnets, including twisted bilayer graphene heterostructures, where quantum geometric quantities can achieve large values. 
\end{abstract} 

\maketitle
Electronic motion in Bloch bands is governed by the bands' dispersion relation and quantum geometry.
The dispersion relation provides an effective kinetic energy for Bloch electrons, and describes classical effects such as acceleration in response to forces. Quantum geometry, on the other hand, as quantified by the Bloch band Berry curvature~\cite{xiao-rmp} and the quantum metric~\cite{provost, berry}, describes the structure of Bloch wave functions. These properties are responsible for a variety of ``non-classical'' effects such as anomalous velocity~\cite{xiao-rmp} and nontrivial modifications to the effective interaction potential between electrons~\cite{yangbo2021}. Quantum geometric effects are particularly pronounced in topological as well as ultra-clean narrow band materials, wherein electronic dispersion is quenched, manifesting in diverse phenomena including quantized transport~\cite{zhang2013}, localization~\cite{marzari97, marzari07}, orbital magnetism~\cite{zhu,repelin}, and superfluidity~\cite{torma}.

\begin{figure}
\includegraphics[width=\columnwidth]{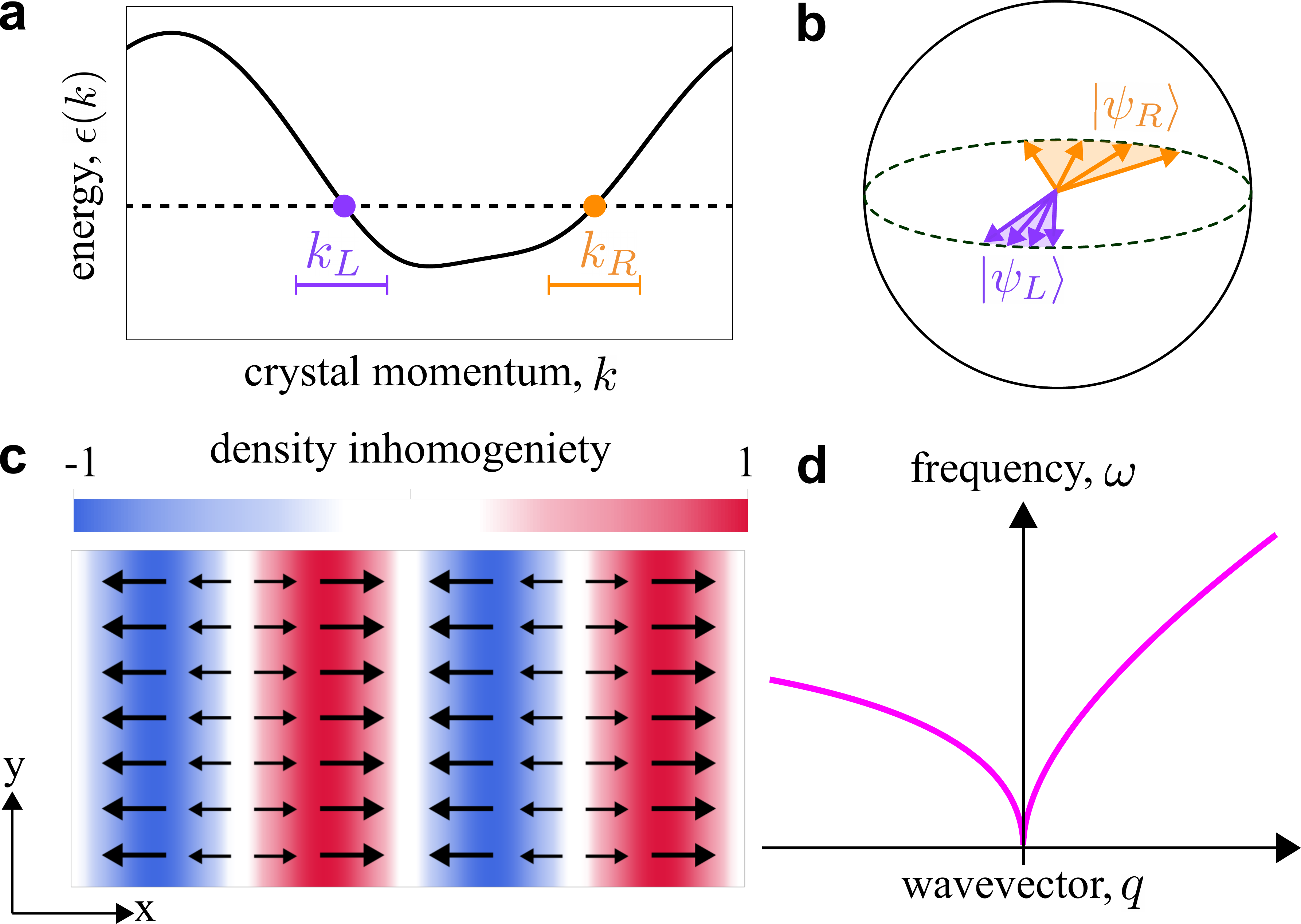}
\caption{When inversion $\mathcal{P}$ and time-reversal $\mathcal{T}$ symmetries are broken, both the (a) electronic dispersion and (b) Bloch wavefunctions are generically asymmetric. Here $k_{L,R}$ denote left and right crystal momentum windows close to the Fermi surface. Despite this asymmetry, spatially uniform linear-response electrical transport is reciprocal. Bulk collective plasmonic modes of the interacting electron liquid, on the other hand, naturally display non-reciprocity when $\mathcal{P}$ and $\mathcal{T}$ are broken. This results from (c) a finite bulk directional current (BDC) (black arrows) which maintains its spatial pattern for both forward and backward propagating plasmon modes (at finite wavevector $\vec q$) thereby producing (d) a non-reciprocal plasmon dispersion.} 
\label{fig1}
\end{figure}

Here we demonstrate the central role of the quantum metric in the collective dynamics of interacting electronic systems~\cite{acsreview}: we show that an asymmetric quantum metric across the Fermi surface in parity-violating (non-centrosymmetric) magnetic metals~\cite{wunderlich2016,tang2016,jungwirthscience,suyang} can drive non-reciprocal propagation of bulk plasmon waves, distinguishing forward and backward propagation. The absence of inversion (parity, $\mathcal{P}$) and time-reversal ($\mathcal{T}$) symmetries in such materials generically leads to a single particle dispersion that is asymmetric with respect to reversal of crystal momentum, $\vec{k}$, see Fig.~\ref{fig1}a. Nonetheless, elementary arguments reveal that spatially uniform linear-response electrical transport remains {\it reciprocal}~\cite{nagaosa}. Instead, as we will describe, non-reciprocity manifests in the collective modes of the interacting electron liquid at nonzero wave vector $\vec{q}$. 

Microscopically, we identify two sources of non-reciprocity: i) a ``classical'' contribution that arises from the asymmetry of the dispersion, and ii) a ``quantum metric'' contribution that arises from the different structure of the system's Bloch wave functions at opposite sides of the Fermi surface, Fig.~\ref{fig1}b. As we will explain, both contributions yield a bulk directional current (BDC) and non-reciprocal plasmon propagation, Fig.~\ref{fig1}c,d. Strikingly, quantum metric plasmon (QMP) non-reciprocity persists even when the single particle dispersion is tuned to be symmetric (or flat), when classical arguments would predict symmetric (reciprocal) behavior. Importantly, for strong interactions (as realized in relatively narrow bands), we find that the quantum metric contribution dominates plasmonic non-reciprocity.

We anticipate a wide range of currently available parity-violating magnets~\cite{wunderlich2016,tang2016,jungwirthscience,suyang} can host bulk non-reciprocal plasmons. Of particular interest are narrow band moir\'e systems which possess parity-violating magnetic states \cite{gordon,young}, as well as exposed surface states that allow direct access via scanning near-field techniques~\cite{franknf,basovnf}. In particular, in twisted bilayer graphene heterostructures close to 3/4 filling, we find strong Coulomb interactions in the narrow bands may enable the quantum metric (dipole) to dominate plasmonic non-reciprocity. Interestingly, non-reciprocal plasmons in twisted bilayer graphene lie in THz range and may play a role in design of miniaturized optical components for THz technology. Another platform of interest is parity violating magnetic systems \cite{jungwirthscience,suyang} (e.g., MnBi$_2$Te$_4$ layers or CuMnAs) where the composite $\mathcal{PT}$ symmetry is preserved (while $\mathcal{P}$ and $\mathcal{T}$ are individually broken). In such systems the Hall effect vanishes, while BDC and intrinsic plasmonic non-reciprocity persists. Additionally, we note that the bulk, intrinsic (i.e.,~passive without driving), and magnetic-field free nature of QMP non-reciprocity distinguishes it from that in other types of chiral plasmons that include edges of Hall metals~\cite{fetter,divincenzo2013,kumar,cbp,reilly},  out-of-equilibrium protocols~\cite{levitov,polini,tobias,cyprian,basov_fizeau}, or magneto-hydrodynamic modes~\cite{sano}.
\\\\

\textit{Semiclassical picture of quantum metric plasmons}\\

To illustrate the origins of QMPs we first analyze the semiclassical collective dynamics of an electron liquid in a slowly varying electric field $\vec E(\vec r, t)$. A full quantum mechanical treatment via the random phase approximation (RPA) follows thereafter. 

Semiclassically, the velocity of an electron with wavevector $\vec k$ is given by~\cite{hughes,dixiaoqm}
\be
\dot{\vec r} (\vec k, t) = \frac{1}{\hbar} \nabla_{\vec k}\Big[ \epsilon (\vec k) +  \delta \epsilon_{\rm QM}(\vec k)\Big] +
\frac{e}{\hbar} \vec E(\vec r, t) \times \boldsymbol{\Omega} (\vec k).
\label{eq:eom}
\ee
Here $\epsilon (\vec k)$ is the Bloch band energy, $\boldsymbol{\Omega} (\vec k)$ is the Berry curvature, and $\delta \epsilon_{\rm QM}(\vec k) =  \tfrac{e}{2} \nabla_{\vec r} \cdot [\mathbf{g} (\vec k) \vec E(\vec r, t)]$, where the rank-2 tensor $\vec{g}(\vec{k})$ is the quantum metric. For the Bloch state $|\psi(\vec{k})\langle$, the quantum metric can be expressed as $g_{ab}(\vec{k})= \textbf{Re}[\langle\partial_{k_a}\psi(\vec{k})|\partial_{k_b}\psi(\vec{k})\rangle] - A_a(\vec{k})A_b(\vec{k})$ where $\vec{A}(\vec{k})$ is the Berry connection. Roman indices $a,b,c$ label Cartesian directions; repeated indices are summed.
The first and last terms of Eq.~(\ref{eq:eom}) are the group velocity [$\vec v_{\vec k} = (1/\hbar)\nabla_{\vec k} \epsilon (\vec k)$] and anomalous velocity, respectively. These contributions arise even under uniform electric fields. In contrast, the second term (a ``quantum metric velocity'' contribution) only appears for non-uniform electric fields and originates from the coupling of spatially inhomogeneous electric fields to the ($\vec k$-dependent part) of the electric quadrupole moment~\cite{hughes,dixiaoqm}. 

Plasmons arise as self-sustained collective density oscillations: $\partial_t \delta n(\vec r, t) + \boldsymbol{\nabla}_{\vec r} \cdot \vec j(\vec r, t) = 0$, where $\delta n(\vec r, t)$ is the deviation of the particle density from equilibrium, and $\vec{j}(\vec{r},t)$ is the particle current density. Decomposing into Fourier harmonics $e^{i(\vec{q} \cdot \vec{r} - \omega t)}$, we find the divergence of the particle current density
\be
i\vec q \cdot \tilde{\vec j}= q_a   \left\{ e q_b\int \dbar \vec k \left[ \frac{i v^a_{\vec k} v^b_{\vec k} }{\omega - \vec q \cdot \vec v_{\vec k}} - i q_c v^a_\vec{k}  g_{bc}(\vec k)\right]\frac{\partial f_0}{\partial \epsilon} \tilde \Phi\right\}, 
\label{eq:current}
\ee
where $\tilde{\mathcal{O}}$ denotes the amplitude of the Fourier harmonic $\mathcal{O}(\vec q, \omega)$. Here $f_0(\epsilon)$ is the Fermi function, and $\Phi (\vec r, t)$ is the electric potential. Here and throughout we use the short hand $\int \dbar \vec k \equiv \int d^d \vec k/(2\pi)^d$ to denote the $d$-dimensional $k$-space integral. The result in Eq.~(\ref{eq:current}) accounts for quadrupolar contributions to the current density and has been simplified by using $\sum_{abc} q_a q_b q_c v^a_{\vec k} g_{bc} (\vec k) =  \sum_{abc} q_a q_b q_c v^b_{\vec k} g_{ac} (\vec k)$.

The current is driven by the plasmon's self-induced electric potential $e\Phi (\vec r, t) = - \int V(\vec r, \vec r') \delta n (\vec r',t) d\vec r' $ generated by $\delta n(\vec r, t)$, where $V(\vec r, \vec r')$ is the Coulomb interaction. In Fourier space, this corresponds to $e\tilde{\Phi}(\vec q, \omega) = -\tilde{V}(|\vec q|) \delta \tilde{n} (\vec q, \omega)$. Solving the continuity equation with this relation and the current density in Eq.~(\ref{eq:current}) produces plasmonic modes.

It is instructive to analyze the behavior of the particle current density in the curly parentheses of Eq.~(\ref{eq:current})~\cite{hall} for forward $e^{i\vec q \cdot \vec r - i \omega t}$ and backward $e^{-i\vec q \cdot \vec r - i \omega t}$ propagating waves. In the absence of the quantum metric and in the long-wavelength limit ($\omega \gg \vec v_{\vec k} \cdot \vec q$), Eq.~(\ref{eq:current}) is dominated by the normal Drude contribution that goes as $\omega^{-1}$. This captures a {\it reciprocal} particle current density that switches sign when the propagation direction is reversed ($\vec q \to - \vec q$). These currents feed the build-up of charge density 
to produce the usual bulk plasmon dispersion relation $\omega (\vec q) =  \omega_0(\vec q)$, with bare plasmon frequency
\be
\omega_0(\vec q) \equiv \sqrt{\tilde{V} ({\vec q})\vec q^T \boldsymbol{\mathcal{D}} \vec q}, \ [\boldsymbol{\mathcal{D}}]_{ab} =  -\int \dbar\vec k \frac{\partial f_0 }{\partial \epsilon} v^a_{\vec k} v^b_{\vec k}, 
\label{eq:bare}
\ee
where $\boldsymbol{\mathcal{D}}$ is the Drude weight. Since the currents that lead to $\omega_0(\vec q)$ are reciprocal, we have $\omega_0(\vec q) = \omega_0(-\vec q)$. 

In contrast, contributions of even powers in $1/\omega$ in the curly parentheses of Eq.~(\ref{eq:current}) are non-reciprocal (e.g., the metric term, $g_{cb} (\vec k)$, as well as other even power contributions in the expansion of the first term). When the propagation direction is reversed, the particle current density pattern described by these contributions {\it maintains} its direction (black arrows, Fig.~\ref{fig1}). We term these contributions bulk directional currents (BDCs); BDCs only arise when both $\mathcal{P}$ and $\mathcal{T}$ symmetry are broken~\cite{symmetry-note}. 

Importantly, BDCs induce non-reciprocal bulk plasmon dispersion relations. To see this explicitly, we first make the simple replacement $\omega - \vec q \cdot \vec v_{\vec k} \approx \omega$ in Eq.~({\ref{eq:current}); see below for a systematic derivation and expansion (in $\vec q$) of the plasmon dispersion. This produces QMPs with a non-reciprocal bulk plasmon dispersion: 
$
\omega(\vec q) = [ \omega_0^2(\vec q)+ \omega_{\rm QM}^2(\vec q)]^{1/2} + \omega_{\rm QM} (\vec q), 
$ 
where the non-reciprocity $\Delta \omega(\vec q) = \omega(\vec q) - \omega(-\vec q)$ reads as 
\be 
\Delta \omega(\vec q) = 2  \omega_{\rm QM} (\vec q), \ \omega_{\rm QM} (\vec q)= \tilde{V}({\vec q})\mathcal{G}^{abc}q_aq_bq_c/2. 
\label{eq:qmplasmons}
\ee
The quantum metric dipole 
\begin{equation}
\mathcal{G}^{abc} = \int \dbar\vec k \,  \frac{\partial f_0}{\partial \epsilon} v_{\vec k}^a g_{bc}(\vec{k})
\label{eq:qmetric} 
\end{equation}
measures the asymmetry of the quantum metric $g_{bc}(\vec{k})$ around the Fermi surface. Crucially, because $\omega_{\rm QM} (\vec q)$ is odd in $\vec q$, the {\it bulk} plasmon dispersion is non-reciprocal: $\omega(\vec q) \neq \omega (- \vec q)$ with a non-reciprocity controlled by the strength of the quantum metric dipole, Eq.~(\ref{eq:qmplasmons}). While in Eq.~(\ref{eq:qmplasmons}) we have focussed on non-reciprocity dominated by $\mathcal{G}^{abc}$, we note that non-reciprocity can also be induced by other types of BDCs so long as $\mathcal{P}$ and $\mathcal{T}$ are broken (see below). Interestingly, as we will see below, QM induced non-reciprocity can persist even in systems with symmetric electronic group velocities, highlighting a non-classical origin of QMP non-reciprocity.
\\\\
\textit{Non-reciprocal bulk plasmons in RPA}\\\\
To obtain bulk plasmon non-reciprocity from a systematic quantum mechanical perspective, we examine screening in a metal, where plasmons appear as zero modes of the dielectric function $\varepsilon(\vec{q},\omega) =  1 - \tilde V(\vec{q}) \Pi(\vec{q},\omega)$. In the random phase approximation (RPA), the polarization function is~\cite{bruus}
\begin{equation}
\label{eq:lindhard}
\Pi(\vec{q},\omega) = \sum_{\alpha,\beta,\vec{k}}\frac{f_0[\epsilon_\alpha(\vec{k}+\vec{q})] - f_0[\epsilon_\beta(\vec{k})]}{\epsilon_\alpha(\vec{k}+\vec{q}) - \epsilon_\beta(\vec{k}) - \hbar\omega} \mathcal{F}_{\alpha\beta}(\vec{k},\vec{q}), 
\end{equation}
where $\mathcal{F}_{\alpha\beta}(\vec{k},\vec{q}) = |\langle\psi_\alpha(\vec{k}+\vec{q})|\psi_\beta(\vec{k})\rangle|^2$ is the coherence factor. We use Greek letters to denote band indices throughout this work. The electronic states $\{|\psi_\alpha(\vec{k})\rangle\}$ have corresponding energies $\{\epsilon_\alpha(\vec{k})\}$. Non-reciprocal plasmons can be obtained as solutions to $\textbf{Re}[\varepsilon(\vec{q},\omega)]=0$, see full numerical implementation below.

To get better intuition about the role of the quantum metric and terms higher order in $\vec q$ in the polarization function, we first analyze the intraband polarization function, focussing on $\alpha=\beta$ in Eq.~(\ref{eq:lindhard}). Expanding in powers of $q = |\vec{q}|$, and using $|\langle\psi ({\vec{k}+\vec{q}})|\psi ({\vec{k}})\rangle|^2 = 1 - g_{ab} (\vec k) q_a q_b - \tfrac{1}{2}\partial_{k_a} g_{bc} (\vec k) q_a q_b q_c + \mathcal{O}(q^4)$ (see \textbf{SI}), gives 
\begin{equation}
\label{eq:lindhardintra}
\Pi_{\rm intra}(\vec{q},\omega) \approx \frac{q_aq_b}{\omega^2} [\boldsymbol{\mathcal{D}}]_{ab} 
+ \frac{q_a q_bq_c}{\omega} \left[ \mathcal{G}^{abc} + \frac{\mathcal{C}^{abc}}{\omega^2} \right]. 
\end{equation} 
Here $C^{abc}$ is a Drude-like weight 
\begin{equation}
\label{eq:drude2}
\mathcal{C}^{abc} = -\int {\dbar\vec k} \,  v_a(\vec k) v_b(\vec{k}) v_c(\vec{k}) \frac{\partial f_0}{\partial \epsilon} 
\end{equation}
that measures the velocity asymmetry of electrons on the Fermi surface. In obtaining Eq.~(\ref{eq:lindhardintra}), we have kept all non-vanishing terms to lowest order up to $\mathcal{O}(q^3)$. We note that just as the quantum metric term $\mathcal{G}^{abc}$ leads to non-reciprocity, the $\mathcal{C}^{abc}$ term (being odd in $q$) is similarly non-reciprocal.
Indeed, both arise under the same symmetry conditions: both $\mathcal{C}$ and $\mathcal{G}$ tensors vanish in the presence of either $\mathcal{T}$ or $\mathcal{P}$ symmetries. We note the origin of $\mathcal{C}^{abc}$, however, is essentially classical, and, as we will see below, vanishes when the electronic dispersion is symmetric; in contrast, $\mathcal{G}$ can persist even when electronic dispersion is symmetric and produce non-reciprocal QMPs. 

Lastly, we note that when both $\mathcal{C}$ and $\mathcal{G}$ are present, for large frequencies ($\omega^2 \gg \mathcal{C}/\mathcal{G}$), quantum metric induced non-reciprocity dominates over that produced by $\mathcal{C}^{abc}$. Indeed, in this regime, approximating the dielectric function with its intraband contribution in Eq.~(\ref{eq:lindhardintra}) produces the non-reciprocal bulk plasmon dispersion in Eq.~(\ref{eq:qmplasmons}). As we will see below, in the presence of strong interactions (e.g., in narrow bands), this produces a regime wherein the quantum metric is dominant. Even as we have concentrated on quantum metric dominated plasmons, we note that BDCs are general (arising from either $\mathcal{C}$ or $\mathcal{G}$ terms), and yield passive, magnetic field free non-reciprocal bulk plasmons. These characteristics in combination are highly attractive for miniaturized optical isolators, and cannot be collectively achieved by present proposals.
\\\\

\textit{Quantum metric plasmons in one-dimension}\\\\
To elucidate the physical nature of QMPs, we now demonstrate their appearance in a simple one-dimensional (1D) model. After using this model to develop a transparent conceptual picture of the physics of QMPs we will turn to examining non-reciprocal bulk plasmons in twisted bilayer graphene heterostructures -- a promising candidate material for their observation.

\begin{figure}
\includegraphics[width=\columnwidth]{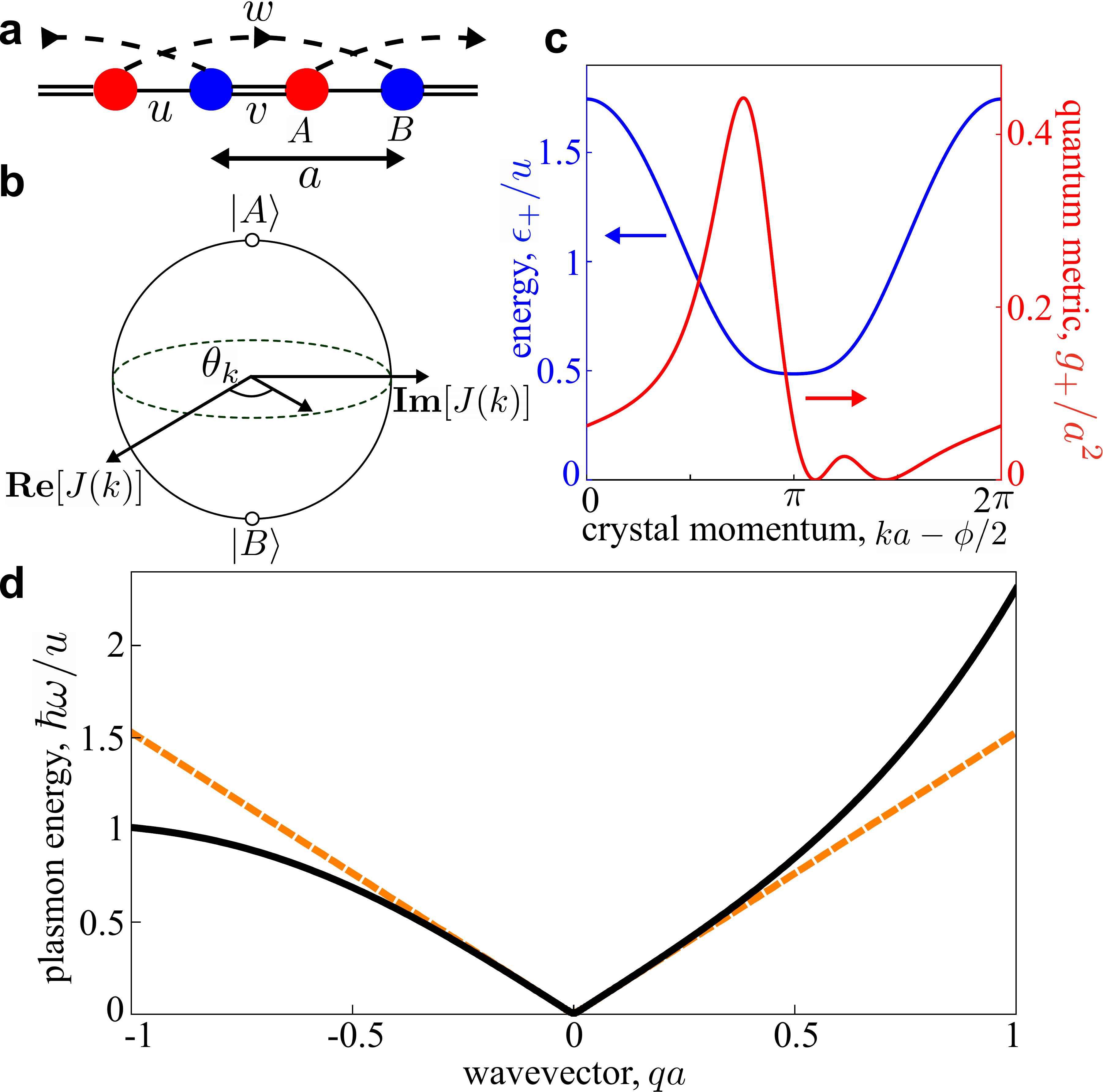}
\caption{(a) 1D magnetic bipartite model with nearest-neighbor hoppings $u$ and $v$, and complex third neighbor hopping $w=|w|e^{i\phi}$. (b) Bloch sphere depicting the phase $\theta_k$ determining the states $|\psi_+(k)\ra$; the rate of change of $\theta_{k}$ determines the quantum metric. (c) Upper band energy [$\epsilon_+(k)$, blue] and the corresponding quantum metric (red), $g_+(k)=[\partial_k \theta_+(k)]^2/4$ for 1D model. (d) Nonreciprocal plasmon dispersion (solid black line) for the electronic Hamiltonian in Eq.~(\ref{eq:H1d}) obtained as the zero modes of the dielectric function; here we have approximated $\Pi (\vec q, \omega)$ by Eq.~(\ref{eq:lindhardintra}). Note that if we set the quantum metric contribution in Eq.~(\ref{eq:lindhardintra}) to 0 (i.e., $\mathcal{G}=0$) [orange dashed line], plasmon non-reciprocity vanishes. Parameters used: $v=|w|=0.4u$ and $\phi=1$; in (c) we have additionally used $\mu=0.6u$ and a contact interaction $\tilde{V}_{1{\rm D}}$ with $ \tilde{V}_0= \tilde{V}_{1{\rm D}}/(ua)= 20$.}
\label{fig2}
\end{figure}

\begin{figure*}
\centerline{\includegraphics[width=\textwidth]{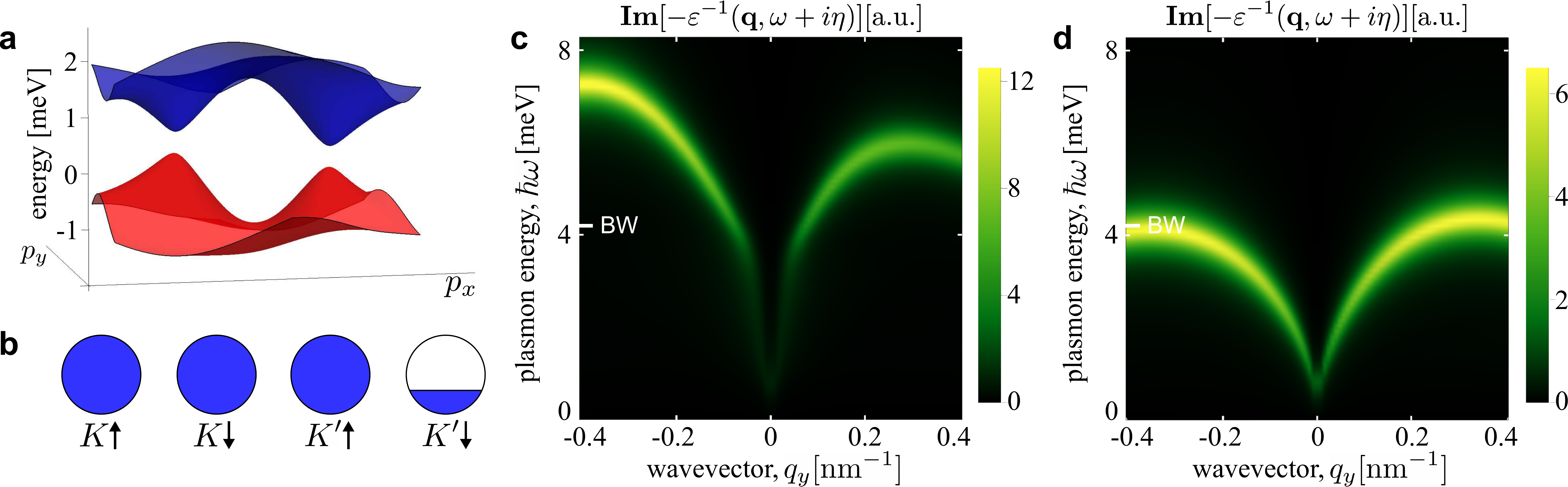}}
\caption{(a) Narrow bands in heterostrained TBG on hBN. Parameters used: hBN staggered potential of 0 and 0.5 meV for top and bottom layers respectively, 0.05\% heterostrain~\cite{bediako}, and interlayer tunneling parameters $t_{\rm AA}=20$ meV and $t_{\rm AB}'=97.5$ meV \cite{frank}. (b) Schematic of spontaneous time reversal symmetry breaking around 3/4 filling displaying valley/spin polarization. (c) RPA loss function $\textbf{Im}[-\varepsilon^{-1}(\vec{q},\omega+i\eta)]$ showing quantum metric non-reciprocal plasmons for strained TBG on hBN at 3/4 filling. Here the chemical potential was set within the conduction flat band at $\mu=1.5$ meV, temperature $T=10$ K, wavevector $\vec{q}=(0,q_y)$, and used backgraound dielectric constant $\kappa=3$ (typical for air/TBG/hBN heterostructures). (d) In contrast to panel (c), when the wavefunction overlaps in Eq.~(\ref{eq:lindhard}) are set to $\mathcal{F}_{\alpha\beta} = \delta_{\alpha\beta}$ (i.e. only classical contributions allowed), plasmonic non-reciprocity is suppressed.
}
\label{fig3}
\end{figure*}

As depicted in Fig.~\ref{fig2}a, we consider a bipartite 1D lattice with uniform on-site energies (set to zero without loss of generality), and intra- and inter-cell nearest-neighbor hopping amplitudes $u$ and $v$, respectively. We additionally include a complex third-neighbor hopping amplitude $w=|w|e^{i\phi}$, with phase convention as indicated by the arrows on the figure. A nonzero value of the phase $\phi$ breaks $\mathcal{T}$ and $\mathcal{P}$ symmetries, as required for the emergence of non-reciprocal plasmons. In the ordered basis $\{\Ket{k, A}, \Ket{k, B}\}$, the system's Bloch Hamiltonian is 
\begin{equation}
\label{eq:H1d} H_{\rm 1D}(k) = \begin{pmatrix}
0 & J^*(k) \\
J(k) & 0
\end{pmatrix}, 
\end{equation} 
with $J(k) = u e^{-ika/2} + v e^{ika/2} + |w| e^{-i(3ka/2-\phi)}$, where $a$ is the lattice constant. The dispersion relations of the system's upper ($+$) and lower ($-$) bands follow from Eq.~(\ref{eq:H1d}): $\epsilon_\pm(k) = \pm|J(k)|$. Note that $H_{\rm 1D}(k)$ has a chiral symmetry associated with the fact that all of its nonzero matrix elements are off-diagonal in sublattice indices.

The quantum metric measures how rapidly the system's Bloch states change direction in Hilbert space as $k$ is varied. For the chiral symmetric model considered here, the system's Bloch states are constrained to lie in the equatorial plane of the Bloch sphere (see Fig.~\ref{fig2}b), with the $+$ state $\Ket{\psi_+(k)}$ oriented along azimuthal angle $\theta(k) = {\rm arg}[J(k)]$. For this case, the quantum metric in band $\pm$ is simply given by $g_\pm(k) = [\partial_k \theta_\pm(k)]^2/4$.

To emphasize the role of the quantum metric on non-reciprocal plasmons we first consider the case of $v=|w|$ and $\phi\neq0$ (Fig.~\ref{fig2}c). In this case, even though $\mathcal{T}$ and $\mathcal{P}$ symmetries are broken, the band energy $\epsilon_+(k) = [(u+2 v{\rm cos}(\phi/2) {\rm cos} \tilde{k})^2 + 4 v^2 {\rm sin}^2 (\phi/2) {\rm cos}^2 \tilde{k}]^{1/2}$ is {\it even} about the band minimum at $\tilde{k} = \pi$, where $\tilde{k} = ka - \phi/2$ (blue curve in Fig.~\ref{fig2}c). As a result, the electronic group velocity is odd about the band minimum, leading to a vanishing $\mathcal{C}$. In contrast, the quantum metric (red curve in Fig.~\ref{fig2}c) remains asymmetric, yielding a finite $\mathcal{G}$. Consequently, the relative shapes of the electronic wave packets at opposite Fermi points can differ significantly~\cite{hughes}. This asymmetry directly leads to non-reciprocal coupling to electromagnetic fields and shows that $\mathcal{G}$ (and as we will now see, QMP non-reciprocity) captures broken $\mathcal{T}$ and $\mathcal{P}$ symmetries as manifested in the structure of the electronic wavefunctions but hidden from the band energies.

To demonstrate the non-reciprocity in QMPs, we obtain the plasmon dispersion relation as zero modes of the dielectric function for the case where the chemical potential is in the upper band; for simplicity, we approximate the polarization function with Eq.~(\ref{eq:lindhardintra}) and use a contact interaction, $\tilde{V}_{1{\rm d}}$~\cite{kane}. Choosing ${V}_0 = \tilde{V}_{1{\rm d}}/ua = 20$  we obtain non-reciprocal QMP dispersion in Fig.~\ref{fig2}d (black). Note that in the absence of the quantum metric [explored by setting $\mathcal{G}=0$ in Eq.~(\ref{eq:lindhardintra})] we find plasmon non-reciprocity vanishes, yielding a reciprocal dispersion (orange curve). This non-reciprocal behavior of QMPs in Fig.~\ref{fig2}d (black) is particularly striking because it emerges from electrons that exhibit a completely symmetric single-particle electronic dispersion. 

We note that the group velocities  become asymmetric for $v \neq |w|$, yielding finite $\mathcal{C}$. Nevertheless, for large contact interactions (dimensionless ratio: $\tilde{V}_0 = \tilde{V}_{1{\rm d}}/ua \gg 1$), the quantum metric dipole continues to provide a dominant contribution to the plasmonic non-reciprocity; for a full discussion see {\bf SI}. For weaker interactions (smaller $\tilde{V}_0$), nonreciprocity is no longer solely dominated by $\mathcal{G}$. Instead, classical contributions from $\mathcal{C}$ also play a sizable role, see {\bf SI}. 
\\\\
\textit{Quantum metric plasmons in twisted bilayer graphene heterostructures}\\\\
We now examine quantum metric plasmons [i.e., non-reciprocal plasmons dominated by $\mathcal{G}_{abc}$] in a promising candidate system: twisted bilayer graphene (TBG) heterostructures. TBG is a particularly attractive venue for QMPs due to its strong Coulomb interactions with typical values that can exceed the bandwidth of the narrow bands~\cite{cyprianundamped}. Additionally, when combined with hBN, TBG heterostructures break inversion symmetry to display large values of quantum geometric quantities~\cite{pacobcd,kaplan,tbgij,swati} in its narrow bands (Fig.~\ref{fig3}a). Importantly, in such heterostructures, a time-reversal symmetry (TRS) broken ferromagnetic phase emerges close to 3/4 filling~\cite{gordon,young}. 

Guided by this, we focus on the plasmonic excitations of TBG/hBN heterostructures in the magnetic phase near 3/4 filling (see schematic in Fig.~\ref{fig3}b). We model the electrons in this system using the continuum model~\cite{koshinoPRX}, including the effect of hBN alignment~\cite{pacobcd} as well as a modest heterostrain~\cite{liangfu} (found in many TBG devices~\cite{pasupathy,yazdani,choi,bediako}). This model produces the narrow bands shown in Fig.~\ref{fig3}a, see {\bf SI} for details.

We numerically identify the plasmonic dispersion by plotting the loss function $\textbf{Im}[-\varepsilon^{-1}(\vec{q},\omega+i\eta)]$ in Fig.~\ref{fig3}c including both intraband ($\alpha = \beta$) as well as interband ($\alpha \neq \beta$) contributions; $\hbar\eta$ is a lifetime parameter \cite{bruus}. Here we have employed an effective 2D Coulomb potential $\tilde{V}^{2\rm{D}}(\vec q) = 2\pi e^2/(\kappa |\vec{q}|)$ with $\kappa = 3$ the background dielectric constant~\cite{cyprianundamped}, and have concentrated on electrons in the narrow bands~\cite{tbg-note}.

Strikingly, a pronounced non-reciprocal TBG plasmon dispersion emerges in Fig.~\ref{fig3}c with plasmon energies that rise above the bandwidth of the narrow bands. Interestingly, the bulk plasmon non-reciprocity is dominated by the quantum metric, thus yielding QMPs. To see this, when the Bloch overlaps are set artificially to $\mathcal{F}_{\alpha\beta}=\delta_{\alpha \beta}$ in Eq.~(\ref{eq:lindhard}) (i.e., only classical terms are allowed), an almost reciprocal plasmon dispersion relation is observed (see Fig.~\ref{fig3}d). The dichotomy between Fig.~\ref{fig3}c (full polarization function) and Fig.~\ref{fig3}d (only classical) highlights the crucial role that the quantum metric plays in the non-reciprocity of TBG plasmons. As expected from Eq.~(\ref{eq:lindhardintra}), quantum metric induced non-reciprocity is expected to be most pronounced at high frequencies (above $\hbar\omega_{\rm TBG} = \hbar\sqrt{\mathcal{C}/\mathcal{G}} \approx 2$ meV).

The quantum metric dipole produces intrinsic non-reciprocal bulk plasmons even when the single particle dispersion is symmetric or flat. This demonstrates the essential role of Bloch band quantum geometric quantities in determining the mesoscopic dynamics of plasmonic collective modes; indeed, this sensitivity of plasmonic modes and the dynamical response to the internal structure of Bloch wavefunctions may be used to probe an asymmetric quantum metric in metals. While we have concenterated on bulk plasmonic modes, quantum metric as well as dispersively induced BDCs can also alter surface modes; we anticipate that together with other quantum geometric quantities this can give rise to a new range of surface plasmons. Additionally, from a technological perspective, BDCs (of both quantum and classical origin) enable new plasmonic non-reciprocities in the bulk via passive and magnetic field free means. Lastly, we note that even as we have focused on plasmonic collective modes, the Bloch wavefunction overlaps that encode the quantum metric dipole are ubiquitous in describing a range electronic responses. As a result, we expect the quantum metric dipole may play important roles in a variety of magneto-chiral effects~\cite{nagaosa,rikken1997,dixiaoqm,iguchi2015,seki2016,seki2019,ando2020}. 
\\\\
\textbf{Acknowledgements}: This work was supported by Singapore MOE Academic Research Fund Tier 3 Grant MOE2018-T3-1-002 and a Nanyang Technological University start-up grant (NTU- SUG).

\newpage
\onecolumngrid
\newpage
\renewcommand{\theequation}{S\arabic{equation}}
\renewcommand{\thefigure}{S\arabic{figure}}
\renewcommand{\thetable}{S\Roman{table}}
\makeatletter
\makeatother
\setcounter{equation}{0}
\setcounter{figure}{0}

\begin{center}
\textbf{Supplementary Information for\\
 ``Quantum plasmonic non-reciprocity in parity-violating magnets"}
\end{center}
\onecolumngrid
\subsection{Identity for expansion of coherence factor} 
Here we summarize a set of identities which we use to expand the coherence factor $|\langle\psi(\vec{k}+\vec{q})|\psi(\vec{k})\rangle|^2$ up to $\mathcal{O}(q^3)$. We begin with the Taylor expand for small $q$: $|\psi(\vec{k}+\vec{q})\rangle = |\psi(\vec{k})\rangle + q_a|\partial_{k_a}\psi(\vec{k})\rangle + (q_a q_b/2)|\partial_{k_a}\partial_{k_b}\psi(\vec{k})\rangle + (q_a q_b q_c/6)|\partial_{k_a}\partial_{k_b}\partial_{k_c}\psi(\vec{k})\rangle$. Collecting terms, the coherence factor takes the form 
\begin{equation}
|\langle\psi(\vec{k}+\vec{q})|\psi(\vec{k})\rangle|^2 = 1 + q_a q_b F_{ab}^{(2)} + q_a q_b q_c F_{abc}^{(3)} + \mathcal{O}(q^4)
\label{seq:coherence}
\end{equation}
where, as in the main text, repeated indices are implicitly summed. The functions $F^{(2,3)}$ can be explicitly written as 
\begin{equation}
\label{seq:f2}
F^{(2)}_{ab} = \frac{1}{2} \left[\langle\partial_{k_a}\partial_{k_b}\psi(\vec{k})|\psi(\vec{k})\rangle + \langle\psi(\vec{k}|\partial_{k_a}\partial_{k_b}\psi(\vec{k}))\rangle\right] + \langle\psi(\vec{k})|\partial_{k_a}\psi(\vec{k})\rangle \langle\partial_{k_b}\psi(\vec{k})|\psi(\vec{k})\rangle
\end{equation}
and 
\begin{multline}
\label{seq:f3}
F^{(3)}_{abc} = \frac{1}{6}\left[ \langle\partial_{k_a}\partial_{k_b}\partial_{k_c}\psi(\vec{k})|\psi(\vec{k})\rangle + \langle\psi(\vec{k})|\partial_{k_a}\partial_{k_b}\partial_{k_c}\psi(\vec{k})\rangle\right] \\
+ \frac{1}{2}\left[\langle\psi(\vec{k})|\partial_{k_a}\psi(\vec{k})\rangle \langle\partial_{k_b}\partial_{k_c}\psi(\vec{k})|\psi(\vec{k})\rangle + \langle\psi(\vec{k})|\partial_{k_a}\partial_{k_b}\psi(\vec{k})\rangle \langle\partial_{k_c}\psi(\vec{k})|\psi(\vec{k})\rangle \right].
\end{multline}

We first examine $F_{ab}^{(2)}$. Recalling the normalization condition $\langle\psi(\vec{k})|\psi(\vec{k})\rangle = 1$ yields $\partial_{k_a}\partial_{k_b}\langle\psi(\vec{k})|\psi(\vec{k})\rangle=0$ which helps us express the first term in Eq. (\ref{seq:f2}) as $-\textbf{Re}[\langle\partial_{k_a}\psi(\vec{k})|\partial_{k_b}\psi(\vec{k})\rangle]$. Similarly, the second term in Eq. (\ref{seq:f2}) can be instantly written in terms of the Berry connection: $\langle\psi(\vec{k})|\partial_{k_a}\psi(\vec{k})\rangle \langle\partial_{k_b}\psi(\vec{k})|\psi(\vec{k})\rangle=A_a(\vec{k}) A_b(\vec{k})$ where $A_a = i\langle\psi(\vec{k})|\partial_{k_a}\psi(\vec{k})\rangle$. Putting these together, enables us to identify $F_{ab}^{(2)}$ in terms of the quantum metric 
\begin{equation}
g_{ab}(\vec{k}) = \textbf{Re}[\langle\partial_{k_a}\psi(\vec{k})|\partial_{k_b}\psi(\vec{k})\rangle] - A_a(\vec{k}) A_b(\vec{k}) = - F_{ab}^{(2)}. 
\label{seq:quantummetric}
\end{equation}

Next, we examine $q_a q_b q_c F_{abc}^{(3)}$. We again use $\langle\psi(\vec{k})|\psi(\vec{k})\rangle = 1$ so that $\partial_{k_a}\partial_{k_b}\partial_{k_c}\langle\psi(\vec{k})|\psi(\vec{k})\rangle=0$ which helps us express the first term in Eq. (\ref{seq:f3}) as: 
\begin{equation}
\label{seq:f31}
\frac{q_a q_b q_c}{6} \left[ \langle\partial_{k_a}\partial_{k_b}\partial_{k_c}\psi(\vec{k})|\psi(\vec{k})\rangle + \langle\psi(\vec{k})|\partial_{k_a}\partial_{k_b}\partial_{k_c}\psi(\vec{k})\rangle\right] = -\frac{q_a q_b q_c}{2} \partial_{k_a}\textbf{Re}[\langle\partial_{k_b}\psi(\vec{k})|\partial_{k_c}\psi(\vec{k})\rangle].
\end{equation}
Here we have repeatedly summed across dummy indices. In the same fashion, the second term in Eq. (\ref{seq:f3}) can be expressed in terms of the derivative of products of Berry connections:
\be
\label{seq:f32}
\frac{q_a q_b q_c}{2}\left[\langle\psi(\vec{k})|\partial_{k_a}\psi(\vec{k})\rangle \langle\partial_{k_b}\partial_{k_c}\psi(\vec{k})|\psi(\vec{k})\rangle + \langle\psi(\vec{k})|\partial_{k_a}\partial_{k_b}\psi(\vec{k})\rangle \langle\partial_{k_c}\psi(\vec{k})|\psi(\vec{k})\rangle\right]
 = \frac{q_a q_b q_c}{2}\partial_{k_a}[A_b(\vec{k}) A_c(\vec{k})]. 
\ee
By combining the identities in Eq. (\ref{seq:f31}) and (\ref{seq:f32}) we get 
\begin{equation}
q_a q_b q_c F^{(3)}_{abc} = -\frac{q_a q_b q_c}{2} \partial_{k_a}\left[\textbf{Re}[\langle\partial_{k_b}\psi(\vec{k})|\partial_{k_c}\psi(\vec{k})\rangle - A_b(\vec{k}) A_c(\vec{k})\right] = -\frac{q_a q_b q_c}{2} \partial_{k_a}g_{bc}(\vec{k}). 
\label{seq:finalf3}
\end{equation}
Applying Eq.~(\ref{seq:finalf3}) and Eq.~(\ref{seq:quantummetric}) into Eq.~(\ref{seq:coherence}) we obtain the identity
\be
|\langle\psi ({\vec{k}+\vec{q}})|\psi ({\vec{k}})\rangle|^2 = 1 - g_{ab} (\vec k) q_a q_b - \tfrac{1}{2}\partial_{k_a} g_{bc} (\vec k) q_a q_b q_c + \mathcal{O}(q^4)
\ee
used in the main text.
 
 \subsection{Non-reciprocal plasmons and dependence on interaction strength}
In the main text, we described how non-reciprocal quantum metric plasmons arise in a 1D magnetic bipartite model for $v = |w|$ where single-particle electronic band structure is symmetric. In such a case, and as highlighted in the main text, contribution from $\mathcal{C}$ vanishes.  
In this section, we consider the more general case of $v \neq |w|$ where the single-particle electronic band structure can become asymmetric so that the contribution of $\mathcal{C}$ no longer vanishes. Additionally, we also highlight the impact of the (contact) interaction strength $\tilde{V}_0$. 

To do so, in Fig.~\ref{figS1}, we display plasmon dispersion relations obtained from the zeroes of the dielectric function and approximating the polarization function with Eq. (\ref{eq:lindhardintra}) in the main text (black curves). Here the parameters for $u,v, |w|$ and $\phi$ are shown in the caption. To highlight the competition between $\mathcal{C}$ and $\mathcal{G}$, in addition to the plasmon dispersion (black curves) we plot guide lines when QM contribution is set to zero (orange dashed curve), i.e. $\mathcal{G}=0$, as well as when the classical velocity asymmetry contribution is set to zero (green dashed curve), i.e. $\mathcal{C}=0$. 

As shown in Fig.~\ref{figS1}, when interaction strength $\tilde{V}_0$ is large, the plasmon dispersion (black curve) adheres closer to the green guideline (as compared to the orange curve). This demonstrates that at large $\tilde{V}_0$, plasmon non-reciprocity is dominated by the quantum metric. In contrast, for intermediate values of $\tilde{V}_0$ the plasmon dispersion (black curve) lies in between orange and green curves; it adheres closer to the orange curve when $\tilde{V}_0$ is small. This demonstrates how $\tilde{V}_0$ controls the competition between quantum metric dominated non-reciprocity (from the $\mathcal{G}$ contribution) prevalent at large interaction strength, and classical velocity asymmetry dominated non-reciprocity (from the $\mathcal{C}$ contribution) prevalent at low interaction strength.

\begin{figure}
\includegraphics[scale=0.155]{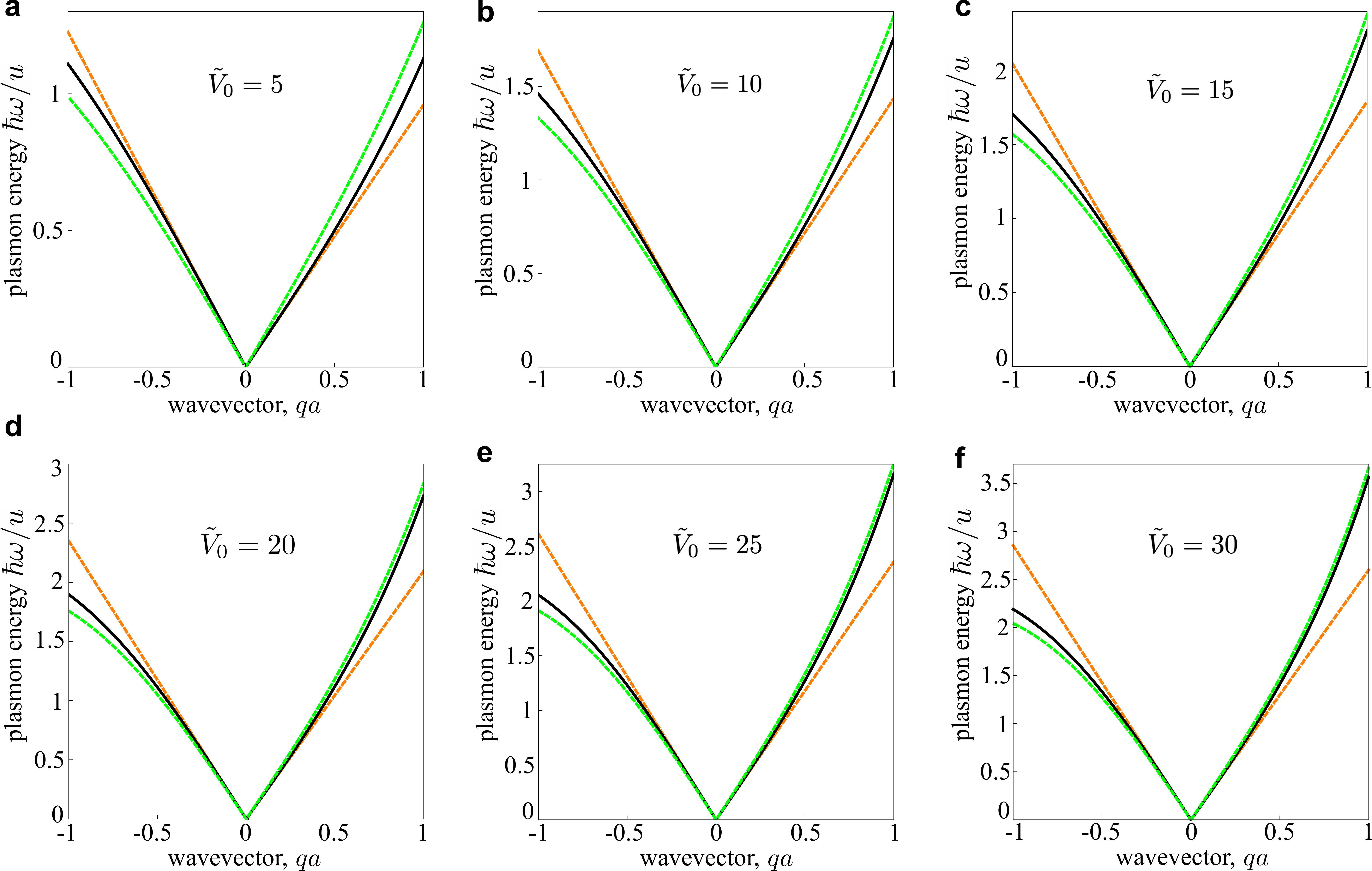}
\caption{Interaction strength dependence of non-reciprocal bulk plasmons. Panel (a)-(f) show plasmon dispersion for different values of $\tilde{V}_0$ as indicated. Here, black solid line shows plasmon dispersion obtained from zeros of the dielectric function approximating $\Pi(\vec q, \omega)$ with Eq.~(\ref{eq:lindhardintra}) of the main text. To highlight the competition between $\mathcal{G}$ and $\mathcal{C}$, we also draw guidelines where quantum metric contribution is set to zero (i.e. $\mathcal{G}=0$) as the orange dashed curve, and where the classical velocity asymmetry contribution is set to zero (i.e. $\mathcal{C}=0$) as the dashed green line. Model parameters: $v=0.5u$, $|w|=0.4u$ and $\phi=1$ along with $\mu=0.8u$.}
\label{figS1}
\end{figure} 
 
 \subsection{Continuum model for strained TBG-hBN heterostructure}
In this section, we detail how we simulated the electronic structure of TBG using the continuum model. For TBG, we define the lattice structure as in Ref. \cite{koshinoPRX}. In each graphene layer the primitive (original) lattice vectors are $\vec{a}_1 = a_{\rm G}(1,0)$ and $\vec{a}_2 = a_{\rm G}(1/2,\sqrt{3}/2)$ with $a_{\rm G}=0.246$ nm being the lattice constant. The corresponding reciprocal space lattice vectors are $\vec{b}_1 = (2\pi/a_{\rm G})(1,-1/\sqrt{3})$ and $\vec{b}_2 = (2\pi/a_{\rm G})(0,2/\sqrt{3})$, and Dirac points are located at $K_\zeta = -\zeta(2\vec{b}_1 +\vec{b}_2)/3$. For a twist angle $\theta$ (accounting for the rotation of layers), the lattice vectors of layer $l$ are given by $\vec{a}_{l,i} = R(\mp\theta/2)\vec{a}_i$, $\mp$ for $l=1,2$ respectively, and $R(\theta)$ represents rotation by an angle $\theta$ about the normal. Also, from $\vec{a}_{l,i}.\vec{b}_{l',j} = 2\pi \delta_{ij}\delta_{ll'}$ we can check that the reciprocal lattice vectors become $\vec{b}_{l,i} = R(\mp\theta/2)\vec{b}_i$ with corresponding Dirac points now located at $\vec{K}_{l,\zeta} =  -\zeta(2\vec{b}_{l,1} +\vec{b}_{l,2})/3$.

At small angles, the slight mismatch of the lattice period between two layers gives rise to long range moir\'e superlattices. The reciprocal lattice vectors for these moir\'e superlattices are given as $\vec{g}_i = \vec{b}_{1,i} - \vec{b}_{2,i}$. The superlattice vectors $\vec{L}$, can then be found using $\vec{g}_i.\vec{L}_j = 2\pi \delta_{ij}$, where $\vec{L}_1$ and $\vec{L}_2$ span the moir\'e unit cell with lattice constant $L = \vec{L}_1 = \vec{L}_2 = a_{\rm G}/[2\sin\theta/2]$. 

Next, when the moir\'e superlattice constant is much longer than the atomic scale, the electronic structure can be described using an effective continuum model for each valley $\zeta=\pm$. The total Hamiltonian is block diagonal in the valley index, and for each valley, the effective Hamiltonian in the continuum model is written in terms of the sublattice and layer basis $(A_1, B_1, A_2, B_2)$ \cite{koshinoPRX}
\begin{equation}
H_\zeta = 	\begin{pmatrix}
H_{1,\zeta}(\vec{p}) & T^\dag_\zeta \\
T_\zeta & H_{2,\zeta}(\vec{p})
\end{pmatrix}
\end{equation}
where $H_{l,\zeta} = -\hbar v_F R(\pm \theta/2)\vec{p}.(\zeta\sigma_x, \sigma_y)$ is the Hamiltonian for each layer with $\hbar v_F/a_{\rm G} = 2135.4$ meV, and 
\begin{equation}
T_\zeta = \begin{pmatrix}
t_{\rm AA} & t'_{\rm AB} \\
t'_{\rm AB} & t_{\rm AA} 
\end{pmatrix} +
\begin{pmatrix}
t_{\rm AA} & t'_{\rm AB}e^{-i\zeta\frac{2\pi}{3}} \\
t'_{\rm AB}e^{i\zeta\frac{2\pi}{3}} & t_{\rm AA} 
\end{pmatrix}e^{i\zeta \vec{g}_1.\vec{r}}
+ 
\begin{pmatrix}
t_{\rm AA} & t'_{\rm AB}e^{i\zeta\frac{2\pi}{3}} \\
t'_{\rm AB}e^{-i\zeta\frac{2\pi}{3}} & t_{\rm AA}
\end{pmatrix}e^{i\zeta (\vec{g}_1+\vec{g}_2).\vec{r}}
\end{equation}
where $\vec{g}_i$ is the reciprocal lattice vector of mBZ. In what follows, we use the tunnelling parameters $t_{\rm AB}=20$ meV and $t'_{\rm AB}=97.5$ meV. These parameters were used to simulate the optical properties of TBG in good agreement with a recent TBG experiment~\cite{frank}. When hBN is aligned with the graphene layers, $C_2$ symmetry is broken modifying the layer Hamiltonians $H_{l,\zeta}$. This can be described by introducing a sublattice staggered potential $\Delta_l$ so that the Hamiltonian for each layer $H_{l,\zeta}(\vec{p}) \rightarrow H_{l,\zeta}(\vec{p})+\Delta_l\sigma_z$ \cite{pacobcd}.

Finally, the presence of a uniaxial heterostrain in TBG of magnitude $\chi$ can be described by the linear strain tensor~\cite{liangfu}
\begin{equation}
\label{eq:strain}
\mathcal{E}_l = \mathcal{F}(l)\chi\begin{pmatrix}
-\cos^2\varphi + \nu \sin^2\varphi & (1+ \nu) \cos\varphi\sin\varphi \\
(1+ \nu) \cos\varphi\sin\varphi & \nu\cos^2\varphi - \sin^2\varphi
\end{pmatrix}
\end{equation}
where $\mathcal{F}(l=1,2)=\mp1/2$, $\nu=0.165$ is the Poisson ratio of graphene and $\varphi$ gives direction of the applied strain. The strain tensor satisfies general transformations in each layer, $\vec{a}_l \rightarrow \vec{a}_l'= [\mathbbm{1}+\mathcal{E}_l]\vec{a}_l$ and $\vec{b}_l \rightarrow \vec{b}_l ' \approx [\mathbbm{1}-\mathcal{E}_l^T]\vec{b}_l$ for real and reciprocal lattice vectors respectively \cite{liangfu}. The strain induced geometric deformations affect the interlayer coupling and further changes the electron motion via gauge field $\vec{A}_l = \sqrt{3}\beta/2a_{\rm G}(\mathcal{E}^{xx}_l+\mathcal{E}^{yy}_l, -2\mathcal{E}^{xy}_l )$, where $\beta = 3.14$. As a result, we have $\vec{p}\rightarrow \vec{p}_{l,\zeta} = [\mathbbm{1}+\mathcal{E}_l^T][\vec{k}-\mathcal{K}_{l,\zeta}]$ with $\mathcal{K}_{l,\zeta} = [\mathbbm{1}-\mathcal{E}_l^T]\vec{K}_{l,\zeta} - \zeta \vec{A}_l$ \cite{liangfu}. 

The effective TBG Hamiltonian modified by the effects of strain and hBN alignment with graphene layers via sublattice staggered potential, can be re-written as 
\begin{equation}
\label{eq:tbghamiltonianstrainandhbn}
\mathcal{H}_\zeta = \begin{pmatrix}
H_{1,\zeta}(\vec{p}_{1,\zeta})+\Delta_1\sigma_z & T_\zeta^\dag \\
T_\zeta & H_{2,\zeta}(\vec{p}_{2,\zeta})+\Delta_2\sigma_z
\end{pmatrix}
\end{equation}
Note that for a given $\vec{q}$ in the mBZ, the 4$\times$4 Hamiltonian in Eq. (\ref{eq:tbghamiltonianstrainandhbn}) is cast into a multiband eigensystem problem as the interlayer coupling leads to hybridisation of the eigenstates at Bloch vectors $\vec{q}$ and $\vec{p}'=\vec{p}+\vec{g}$, where $\vec{g} = m_1\vec{g}_1 +m_2\vec{g}_2$ and $m_{1,2}\in\mathbb{Z}$ \cite{koshinoPRX}. We truncate the size of the matrix by defining a circular cut-off $|\vec{p}-\vec{p}'|<4|\vec{g}_1|$ \cite{koshinoPRX}. For a given Bloch vector $\vec{p}$, this gives us 61 sites in reciprocal space, and a corresponding matrix of size $244 \times 244$ which is then diagonalized to obtain eigenvalues and eigenvectors. These eigenvalues and eigenvectors are then used to solve for polarization function in Eq.~(\ref{eq:lindhard}) on a grid of $400\times 400$ points.

\subsection{Numerical Evaluation of plasmon dispersion in TBG}
The polarization function in Eq. (\ref{eq:lindhard}) is calculated numerically for TBG on hBN with a heterostrain of $\chi=0.05\%$ and hBN staggered potential due to alignment between graphene and hBN layers as $\Delta_1=0$ and $\Delta_2=0.5$ meV. The $k$-space integral over the mBZ was performed as a Riemann sum on a grid of 400$\times$400 points with $\hbar\eta=0.3$ meV for each value of $\vec{q}$ and $\omega$. We include all intra- and inter-band contributions between the flat bands, and neglect the effect of remote bands as our parameters the flat and remote bands separation is nearly 80 meV which is much larger than the energy of plasmons of our interest [see fig. \ref{figS2}]. The corresponding dielectric function is obtained as: 
$\varepsilon(\vec{q},\omega+i\eta) = 1 - \frac{2\pi e^2}{\kappa|\vec{q}|} \Pi(\vec{q},\omega+i\eta)$ which is finally inverted to get the loss function. Here we have used the effective 2D Coulomb potential $\tilde{V}^{2\rm{D}}(\vec q) = 2\pi e^2/(\kappa |\vec{q}|)$ with $\kappa =3$ the background dielectric constant. The plasmons appear as highly peaked values in the loss function as shown in the main text. 

\begin{figure}
\includegraphics[scale=0.2]{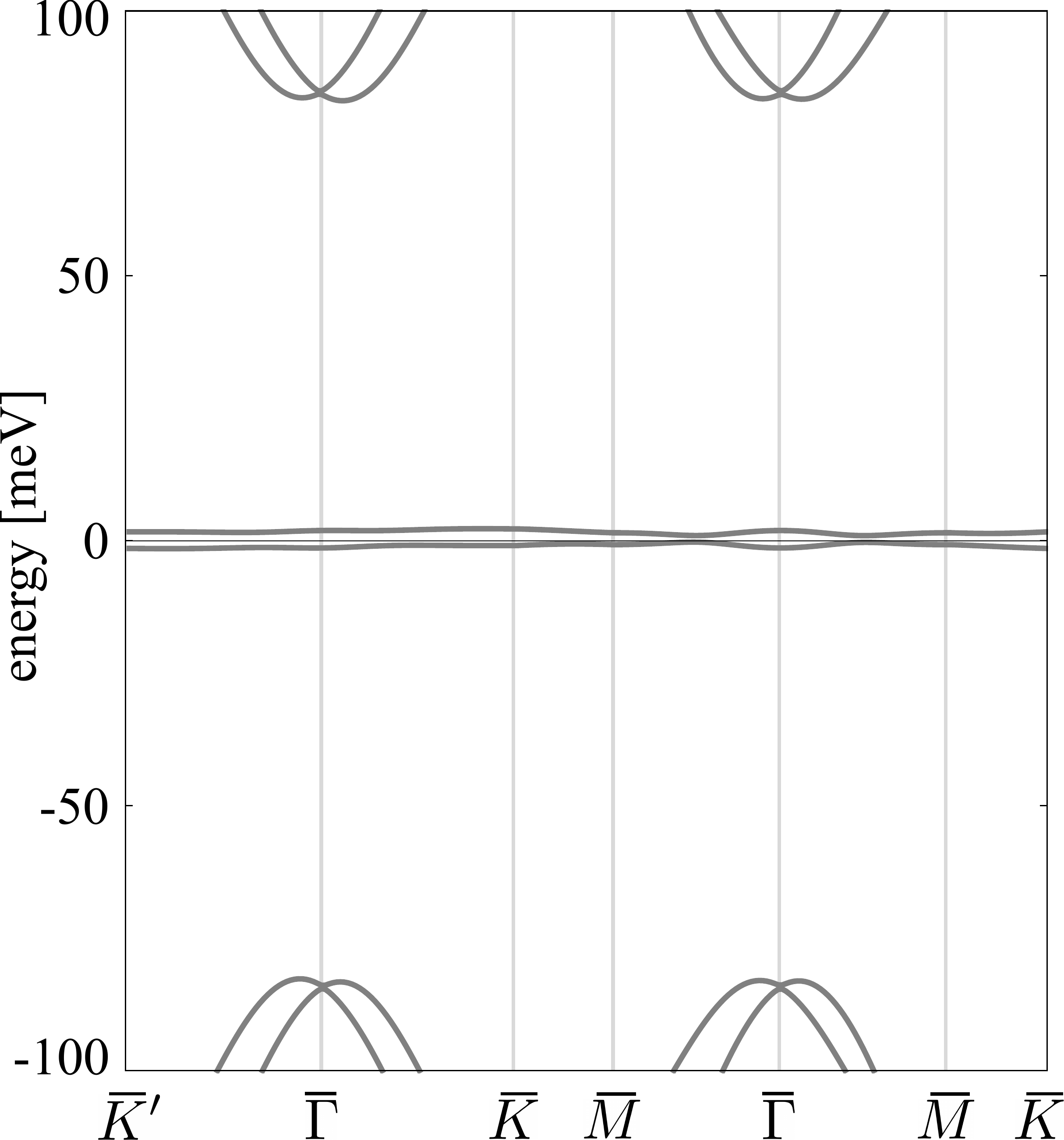}
\caption{TBG band structure along the specified path in mBZ for interlayer couplings $t_{\rm AA}=20$ meV, $t'_{\rm AB}=97.5$ meV with heterostrain of $\chi=0.05\%$. We take sublattice symmetry breaking potentials $\Delta_1=0$ and $\Delta_2=0.5$ meV.}
\label{figS2}
\end{figure}


\begin{thebibliography}{99}

\bibitem{xiao-rmp} Xiao, D.; Chang, M.-C.; Niu, Q. Berry phase effects on electronic properties. \textit{Rev. of Mod. Phys.} \textbf{2010}, 82, 1959.

\bibitem{provost} Provost, J. P.; Vallee, G. Riemannian structure on manifolds of quantum states. \textit{Commun. Math. Phys}. \textbf{1980}, 76, 289.

\bibitem{berry} Berry, M. V. The quantum phase, five years after, in \textit{Geometric Phases in Physics} (World Scientific, Singapore, 1989), pp. 7–28.

\bibitem{yangbo2021} Wang, J.; Cano, J.; Millis, A. J.; Liu, Z.; Yang, B. Exact Landau Level Description of Geometry and Interaction in a Flatband. \textit{Phys. Rev. Lett.} {\bf 2021}, 127, 246403.

\bibitem{zhang2013} Chang, C.-Z,; Zhang, J,; Feng, X.; Shen, J.; Zhang, Z.; Guo, M.; Li, K.; Ou, Y.; Wei, P.; Ji, Z.-Q.; Feng, Y.; Ji, S.-H.; Chen, X.; Jia, J.; Dai, X.; Fang, Z.; Zhang, S.-C.; He, K.; Wang, Y.; Lu, L.; Ma, X.-C.; Xu, Q.-K. Experimental observation of the quantum anomalous Hall effect in a magnetic topological insulator. \textit{Science} \textbf{2013}, 340, 167.

\bibitem{marzari97} Marzari, N.; Vanderbilt, D. Maximally localized generalized Wannier functions for composite energy bands. \textit{Phys. Rev. B} \textbf{1997}, 56, 12847.

\bibitem{marzari07} Brouder, C.; Panati, G.; Calandra, M.; Mourougane, C.; Marzari, N. Exponential localization of Wannier functions in insulators. \textit{Phys. Rev. Lett.} \textbf{2007}, 98, 046402.

\bibitem{repelin} Repellin, C.; Dong, Z.; Zhang, Y. H.; Senthil, T. Ferromagnetism in narrow bands of moir\'e superlattices. \textit{Phys. Rev. Lett.} \textbf{2020}, 124, 187601.

\bibitem{zhu} Zhu, J.; Su, K.-J.; MacDonald, A. H. Voltage-controlled magnetic reversal in orbital Chern insulators. \textit{Phys. Rev. Lett.} \textbf{2020}, 125, 227702.

\bibitem{torma} Peotta, S.; T\"orma, P. Superfluidity in topologically non-trivial flat bands. \textit{Nat. Comm.} \textbf{2015}, 6, 8944.

\bibitem{acsreview} Reserbat-Plantey, A.; Epstein, I.; Torre, I.; Costa, A.T.; Gon\c calves, P. A. D.; Asger Mortensen, N.; Polini, M.; Song, J. C. W.; Peres, N. M. R.; Koppens, F. H. L. Quantum Nanophotonics in Two-Dimensional Materials. \textit{ACS Photonics} \textbf{2021}, 8, 85.

\bibitem{wunderlich2016} Jungwirth, T.; Marti, X.; Wadley, P.; Wunderlich, J. Antiferromagnetic spintronics. \textit{Nat. Nano.} \textbf{2016}, 11, 231.

\bibitem{jungwirthscience} Wadley, P.; Howells, B.; Elezny, J.; Andrews, C.; Hills, V.; Campion, R. P.; Novak, V.; Olejnik, K.; Maccherozzi, F.; Dhesi, S. S.; Martin, S. Y.; Wanger, T.; Wunderlich, J.; Freimuth, F.; Mokrousov, Y.; Kune\"s, J.; Chauhan, J. S.; Grzybowski, M. J.; Rushforth, A. W.; Edmonds, K. W.; Gallagher, B. L.; Jungwirth, T. Electrical switching of an antiferromagnet. \textit{Science} \textbf{2016}, 351, 587.

\bibitem{tang2016} Tang, P.; Zhou, Q.; Xu, G.; Zhang, S.-C. Dirac fermions in an antiferromagnetic semimetal. \textit{Nat. Phys.} \textbf{2016}, 12, 1100.

\bibitem{suyang} Gao, A.; Liu, Y.-F.; Hu, C.; Qiu, J.-X.; Tzschaschel, C.; Ghosh, B.; Ho, S.-C.; B\'erub\'e, D.; Chen, R.; Sun, H.; Zhang, Z.; Zhang, X.-Y.; Wang, Y.-X; Wang, N.; Huang, Z.; Felser, C.; Agarwal, A.; Ding, T.; Tien, H.-J.; Akey, A.; Gardener, J.; Singh, B.; Watanabe, K.; Taniguchi, T.; Burch, K. S.; Bell, D. C.; Zhou, B. B.; Gao, W.; Lu, H.-Z.; Bansil, A.; Lin, H.; Chang, T. R.; Fu, L.; Ma, Q.; Ni, N.; Xu, S.-Y.; Layer Hall effect in a 2D topological axion antiferromagnet. \textit{Nature (London)} \textbf{2021}, 595, 521.

\bibitem{nagaosa} Tokura, Y.; Nagaosa, N. Nonreciprocal responses from non-centrosymmetric quantum materials. \textit{Nat. Comm.} \textbf{2018}, 9, 3740.

\bibitem{gordon} Sharpe, A. L.; Fox, E. J.; Barnard, A. W.; Finney, J.; Watanabe, K.; Taniguchi, T.; Kastner, M. A.; Goldhaber-Gordon, D. Emergent ferromagnetism near three-quarters filling in twisted bilayer graphene. \textit{Science} \textbf{2019}, 365, 605.

\bibitem{young} Serlin, M.; Tschirhart, C. L.; Polshyn, H.; Zhang, Y.; Zhu, J.; Watanabe, K.; Taniguchi, T.; Balents, L.; and Young, A. F. Intrinsic quantized anomalous Hall effect in a moir\'e heterostructure. \textit{Science} \textbf{2020}, 367, 900.

\bibitem{franknf} Chen, J.; Badioli, M.; Alonso-Gonz\'aez, P.; Thongrattanasiri, S.; Huth, F.; Osmond, J.; Spasenovi\c', M.; Centeno, A.; Pesquera, A.; Godignon, P.; Elorza, A Z.; Camara, N.; de Abajo, F. J. G.; Hillenbrand, R.; Koppens, F. H. L. Optical nano-imaging of gate-tunable graphene plasmons. \textit{Nature} \textbf{2012}, 487, 77-81.

\bibitem{basovnf} Fei, Z.; Rodin, A.; Andreev, G. O.; Bao, W.; McLeod, A. S.; Wagner, M.; Zhang, L. M.; Zhao, Z.; Thiemens, M.; Dominguez, G.; Fogler, M. M.; Castro Neto, A. H.; Lau, C. N.; Keilmann, F.; Basov, D. N. Gate-tuning of graphene plasmons revealed by infrared nano-imaging. \textit{Nature} \textbf{2012}, 487, 82.

\bibitem{fetter} Mast, D. B.; Dahm, A. J.; Fetter, A. L. Observation of bulk and edge magnetoplasmons in a two-dimensional electron fluid \textit{Phys. Rev. Lett}. \textbf{1985}, 54, 1706.

\bibitem{divincenzo2013} Viola, G.; DiVincenzo, D. P. Hall effect gyrators and circulators. \textit{Phys. Rev. X} \textbf{2104}, 4, 021019.

\bibitem{cbp} Song, J. C. W.; Rudner, M. Chiral plasmons without magnetic field. \textit{Proc. Nat. Acad. Sci. U.S.A.} \textbf{2016}, 113, 4658.

\bibitem{kumar} Kumar, A.; Nemilentsau, A.; Fung, K. H.; Hanson, G.; Fang, N. X.; Low, T. Chiral plasmon in gapped Dirac systems. \textit{Phys. Rev. B} \textbf{2016}, 93, 041413(R).

\bibitem{reilly} Mahoney, A. C.; Colles, J. I.; Peeters, L.; Pauka, S. J.; Fox, E. J.; Kou, X.; Pan, L.; Wang, K. L.; Goldhaber-Gordon, D.; and Reily, D. J. Zero-field edge plasmons in a magnetic topological insulator. \textit{Nat. Comm.} \textbf{2017}, 8, 1836.

\bibitem{levitov} Borgnia, D. S.; Phan, T. V.; Levitov, L. S. Quasi-relativistic Doppler Effect and non-reciprocal plasmons in graphene. \textit{arXiv:1512.09044} \textbf{2015}.

\bibitem{tobias} Sabbaghi, M.; Lee, H.-W.; Stauber, T.; Kim, K. S. Drift-induced modifications to the dynamical polarization of graphene. \textit{Phys. Rev. B} \textbf{2015}, 92, 195429.

\bibitem{polini} Duppen, B. V.; Tomadin, A.; Grigorenko, A. N.; Polini, M. Current-induced birefringent absorption and non-reciprocal plasmons in graphene. \textit{2D Materials} \textbf{2016}, 3, 015011.

\bibitem{cyprian} Papaj, M.; Lewandowski, C. Plasmonic non-reciprocity driven by band hybridization in moir\'e materials. \textit{Phys. Rev. Lett.} \textbf{2020}, 125, 066801.

\bibitem{basov_fizeau} Dong, Y.; Xiong, L.; Phinney, I. Y.; Sun, Z.; Jing, R.; McLeod, A. S.; Zhang,S .; Liu, S.; Ruta, F. L.; Gao, H.; Dong, Z.; Pan, R.; Edgar, J. H.; Jarillo-Herrero, P.; Levitov, L. S.; Millis, A. J.; Fogler, M. M.; Bandurin, D. A.; and Basov, D. N. Fizeau drag in graphene plasmonics. \textit{Nature (London)} \textbf{2021}, 594, 513.

\bibitem{sano} Sano, R.; Toshio, R.; Kawakami, N. Nonreciprocal electron hydrodynamics under magnetic fields: Applications to non-reciprocal surface magnetoplasmons. \textit{Phys. Rev. B} \textbf{2021}, 104, L241106.

\bibitem{hughes} Lapa, M. F.; Hughes, T. L. Semiclassical wave packet dynamics in nonuniform electric fields. \textit{Phys. Rev. B} \textbf{2019}, 99, 121111(R).

\bibitem{dixiaoqm} Gao, Y.; Xiao, D. Nonreciprocal directional dichroism induced by the quantum metric dipole. \textit{Phys. Rev. Lett.} \textbf{2019}, 122, 227402.

\bibitem{hall} Note that Hall components of the particle current density do not contribute to Eq.~(\ref{eq:current}) since they are divergence free.

\bibitem{symmetry-note} Note that with either $\mathcal{P}$ or $\mathcal{T}$ symmetries, $\vec v_{\vec k} = -\vec v_{-\vec k}$ and $g_{\rm ab}(\vec k) = g_{\rm ab}(-\vec k)$.

\bibitem{bruus} Bruus, H.; Flensberg, K. \textit{Many-body Quantum Theory in Condensed Matter Physics} (Oxford University Press, 2004).

\bibitem{kane} Kane, C.; Balents, L.; Fisher, M. P. A. Coulomb interactions and mesoscopic effects in carbon nanotubes \textit{Phys. Rev. Lett.} \textbf{1997}, 79, 5086.

\bibitem{cyprianundamped} Lewandowski, C.; Levitov, L. S. Intrinsically undamped plasmon modes in narrow electron bands. \textit{Proc. Nat. Acad. Sci. U.S.A.} \textbf{2019}, 116, 20869–20874.

\bibitem{pacobcd} Pantale\'on, P. A.; Low, T.; Guinea, F. Tunable large Berry dipole in strained twisted bilayer graphene. \textit{Phys. Rev. B} \textbf{2021}, 103, 205403.

\bibitem{tbgij} Arora, A.; Kong, J. F.; Song, J. C. W. Strain-induced large injection current in twisted bilayer graphene. \textit{Phys. Rev. B} \textbf{2021}, 104, L241404.

\bibitem{swati} Chaudhary, S.; Lewandowski, C.; Refael, G. Shift-current response as a probe of quantum geometry and electron-electron interactions in twisted bilayer graphene. \textit{Phys. Rev. Res.} \textbf{2022}, 4, 013164 (2022).

\bibitem{kaplan} Kaplan, D.; Holder, T.; Yan, B. Twisted photovoltaics at terahertz frequencies from momentum shift current. \textit{Phys. Rev. Res.} \textbf{2022}, 4, 013209.

\bibitem{koshinoPRX} Koshino, M.; Yuan, N. F. Q.; Koretsune, T.; Ochi, M.; Kuroki, K.; Fu, L. Maximally localized Wannier orbitals and the extended Hubbard model for twisted bilayer graphene. \textbf{2018}, 8, 031087.

\bibitem{liangfu} Bi, Z.; Yuan, N. F. Q.; Fu, L. Designing flat bands by strain. \textit{Phys. Rev. B} \textbf{2019}, 100, 035448 (2019).

\bibitem{pasupathy} Kerelsky, A.; McGilly, L. J.; Kennes, D. M.; Xian, L.; Yankowitz, M.; Chen, S.; Watanabe, K.; Taniguchi, T.; Hone, J.; Dean, C.; Rubio, A.; Pasupathy, A. N. Maximized electron interactions at the magic angle in twisted bilayer graphene. \textit{Nature (London)} \textbf{2019}, 572, 95.

\bibitem{yazdani} Xie, Y.; Lian, B.; J\a"ck, B.; Liu, X.; Chiu, C.-L.; Watanabe, K.; Taniguchi, T.; Bernevig, B. A.; Yazdani, A. Spectroscopic signatures of many-body correlations in magic-angle twisted bilayer graphene. \textit{Nature (London)} \textbf{2019}, 572, 101.

\bibitem{choi} Choi, Y.; Kemmer, J.; Peng, Y.; Thomson, A.; Arora, H.; Polski, R.; Zhang, Y.; Ren, H.; Alicea, J.; Refael, G.; von Oppen, F.; Watanabe, K.; Taniguchi, T.; Nadj-Perge, S. Electronic correlations in twisted bi- layer graphene near the magic angle. \textit{Nat. Phys.} \textbf{2019}, 15, 1174.

\bibitem{bediako} Kazmierczak, N. P.; Van Winkle, M.; Ophus, C.; Bustillo, K. C.; Carr, S.; Brown, H. G.; Ciston, J.; Taniguchi, T.; Watanabe, K.; Bediako, D. K. Strain fields in twisted bilayer graphene. \textit{Nat. Mater.} \textbf{2021}, 20, 956.

\bibitem{tbg-note} For the TBG parameters we used, excitations from remote bands have energies far larger than the plasmon energies considered. 

\bibitem{frank} Hesp, N. C. H.; Torre, I.; Rodan-Legrain, D.; Novelli, P.; Cao, Y.; Carr, S.; Fang, S.; Stepanov, P.; Barcons-Ruiz, D.; Herzig-Sheinfux, H.; Watanabe, K.; Taniguchi, T.; Efetov,  D. K.; Kaxiras, E.; Jarillo-Herrero, P.; Polini, M.; Koppens, F. H. L. Observation of interband collective
excitations in twisted bilayer graphene. \textit{Nat. Phys.} \textbf{2021}, 17, 1162-1168.

\bibitem{rikken1997} Rikken, G. L. J.; Raupach, E. Observation of magneto-chiral dichroism \textit{Nature (London)} \textbf{1997}, 390, 493.

\bibitem{iguchi2015} Iguchi, Y.; Uemura, S.; Ueno, K.; Onose, Y. Nonreciprocal magnon propagation in a noncentrosymmetric ferromagnet LiFe$_5$O$_8$. \textit{Phys. Rev. B} \textbf{2015}, 92, 184419.

\bibitem{seki2016} Seki, S.; Okamura, Y.; Kondou, K.; Shibata, K.; Kubota, M.; Takagi, R.; Kagawa, F.; Kawasaki, M.; Tatara, G.; Otani, Y.; Tokura, Y. Magnetochiral nonreciprocity of volume spin wave propagation in chiral-lattice ferromagnets. \textit{Phys. Rev. B} \textbf{2016}, 83, 235131.

\bibitem{seki2019} Nomura, T.; Zhang, X.-X.; Zherlitsyn, S.; Wosnitza, J.; Tokura, Y.; Nagaosa, N.; Seki, S. Phonon magnetochiral effect. \textit{Phys. Rev. Lett.} \textbf{2019}, 122, 145901.

\bibitem{ando2020} Ando, F.; Miyasaka, Y.; Li, T.; Ishizuka, J.; Arakawa, T.; Shiota, Y.; Moriyama, T.; Yanase, Y.; Ono, T. Observation of superconducting diode effect.\textit{Nature (London)} \textbf{2020}, 584, 373-376.

\end{thebibliography}
\end{document}